\documentclass{aastex6}

\fullcollaborationName{}

\begin{document}


\title{Inverse FIP Effects in Giant Solar Flares Found from Earth X-Ray Albedo with Suzaku/XIS}


\author{Satoru Katsuda\altaffilmark{1}, Masanori Ohno\altaffilmark{2,3,4,5}, Koji Mori\altaffilmark{6}, Tatsuhiko Beppu\altaffilmark{6}, Yoshiaki Kanemaru\altaffilmark{6}, Makoto S. Tashiro\altaffilmark{7,1}, Yukikatsu Terada\altaffilmark{1,7}, Kosuke Sato\altaffilmark{1}, Kae Morita\altaffilmark{1}, Hikari Sagara\altaffilmark{1}, Futa Ogawa\altaffilmark{1}, Haruya Takahashi\altaffilmark{1}, Hiroshi Murakami\altaffilmark{8}, Masayoshi Nobukawa\altaffilmark{9}, Hiroshi Tsunemi\altaffilmark{10}, Kiyoshi Hayashida\altaffilmark{10,11,7}, Hironori Matsumoto\altaffilmark{10,11}, Hirofumi Noda\altaffilmark{10,11}, Hiroshi Nakajima\altaffilmark{12}, Yuichiro Ezoe\altaffilmark{13}, Yohko Tsuboi\altaffilmark{14}, Yoshitomo Maeda\altaffilmark{7}, Takaaki Yokoyama\altaffilmark{15}, and Noriyuki Narukage\altaffilmark{16}}

\altaffiltext{1}{Graduate School of Science and Engineering, Saitama University, 255 Shimo-Ohkubo, Sakura, Saitama 338-8570, Japan}
\altaffiltext{2}{Institute of Physics, E\"otv\"os University, P\'azm\'any P\'eter s\'et\'any 1/A, Budapest, 1117, Hungary}
\altaffiltext{3}{MTA-E\"otv\"os University Lend\"ulet Hot Universe and Astrophysics Research Group, Budapest, Hungary}
\altaffiltext{4}{Konkoly Observatory of the Hungarian Academy of Sciences, Konkoly-Thege ut 15-17, Budapest, 1121, Hungary}
\altaffiltext{5}{School of Science, Hiroshima University, 1-3-1, Kagamiyama, Higashi-Hiroshima, Hiroshima 739-8526, Japan}
\altaffiltext{6}{Department of Applied Physics and Electronic Engineering, University of Miyazaki, 1-1, Gakuen Kibanadai-nishi, Miyazaki 889-2192, Japan}
\altaffiltext{7}{Japan Aerospace Exploration Agency, Institute of Space and Astronautical Science, 3-1-1 Yoshino-dai, Chuo-ku, Sagamihara, Kanagawa 252-5210, Japan}
\altaffiltext{8}{Faculty of Liberal Arts, Tohoku Gakuin University, 2-1-1 Tenjinzawa, Izumi-ku, Sendai, Miyagi 981-3193, Japan}
\altaffiltext{9}{Department of Teacher Training and School Education, Nara University of Education, Takabatake-cho, Nara, Nara 630-8528, Japan}
\altaffiltext{10}{Department of Earth and Space Science, Osaka University, 1-1 Machikaneyama-cho, Toyonaka, Osaka 560-0043, Japan}
\altaffiltext{11}{Project Research Center for Fundamental Sciences, Osaka University, 1-1 Machikaneyama-cho, Toyonaka, Osaka 560-0043, Japan}
\altaffiltext{12}{College of Science and Engineering, Kanto Gakuin University, 1-50-1 Mutsuurahigashi, Kanazawa-ku, Yokohama, Kanagawa 236-8501, Japan}
\altaffiltext{13}{Department of Physics, Tokyo Metropolitan University, 1-1 Minami-Osawa, Hachioji, Tokyo 192-0397, Japan}
\altaffiltext{14}{Department of Physics, Chuo University, 1-13-27 Kasuga, Bunkyo, Tokyo 112-8551, Japan}
\altaffiltext{15}{Department of Earth and Planetary Science, The University of Tokyo, 7-3-1 Hongo, Bunkyo-ku, Tokyo 113-0033, Japan}
\altaffiltext{16}{National Astronomical Observatory of Japan, National Institutes of Natural Science, 2-21-1 Osawa, Mitaka, Tokyo 181-8588, Japan}


\begin{abstract}

We report X-ray spectroscopic results for four giant solar flares occurred on 2005 September 7 (X17.0), 2005 September 8 (X5.4), 2005 September 9 (X6.2), and 2006 December 5 (X9.0), obtained from Earth albedo data with the X-ray imaging spectrometer (XIS) onboard Suzaku.  The good energy resolution of the XIS (FWHM$\sim$100\,eV) enables us to separate a number of line-like features and detect the underlying continuum emission.  These features include Si He$\alpha$, Si Ly$\alpha$, S He$\alpha$, S Ly$\alpha$, Ar He$\alpha$, and Ca He$\alpha$ originating from solar flares as well as fluorescent Ar K$\alpha$ and Ar K$\beta$ from the Earth atmosphere.  Absolute elemental abundances (X/H) averaged over the four flares are obtained to be $\sim$2.0 (Ca), $\sim$0.7 (Si), $\sim$0.3 (S), and $\sim$0.9 (Ar) at around flare peaks.  This abundance pattern is similar to those of active stars' coronae showing inverse first ionization potential (i-FIP) effects, i.e., elemental abundances decrease with decreasing FIP with a turnover at the low end of FIP.  The abundances are almost constant during the flares, with an exception of Si which increases by a factor of $\sim$2 in the decay phase.  The evolution of the Si abundance is consistent with a picture that the i-FIP plasma originates from the chromosphere evaporation and then mixes with the surrounding low-FIP biased materials.  Flare-to-flare abundance varied by a factor of 2, agreeing with past observations of solar flares.  Finally, we emphasize that Earth albedo data acquired by X-ray astronomy satellites like Suzaku and XRISM can significantly contribute to studies of solar physics.  
\end{abstract}




\section{Introduction} \label{sec:intro}

The plasma composition is one of the keys to understanding the physical processes associated with mechanisms to transport, heat, and accelerate the plasma in the solar atmosphere.  In the last half of the 20th century, it has been revealed that the elemental composition of the solar corona is different from that of the underlying photosphere by extreme-ultraviolet and X-ray spectroscopy of the solar corona as well as in-situ measurements of solar-energetic particles and solar winds.  The discrepancy is understood to be caused by a dependence on the first ionization potential (FIP) of the element \citep[e.g.,][]{Meyer1985,Schmelz2012,Reames2018}.  In the corona, elements with low FIP ($\lesssim$ 10\,eV) are enhanced by a factor of 2--4 comapred with the photospheric abundances \citep[e.g.,][]{Feldman1992,Dennis2015}.\footnote{However, it has been still debated whether low-FIP elements are enhanced or high-FIP elements are depleted relative to the photospheric values \citep[e.g.,][]{Veck1981,Raymond1997}.}  This is widely known as the ``FIP effect", but its mechanism has been one of the most enduring mysteries of solar physics, besides the well-known coronal heating problem.

It has been shown that coronae of inactive stars exhibit similar solar-like FIP effects \citep{Drake1997,Laming1999}.  In contrast, later spectral types with stronger magnetic fields show an ``inverse FIP (i-FIP) effect", i.e., low-FIP elements are under-abundant relative to high-FIP elements \citep[e.g.,][]{Brinkman2001,Gudel2001,Audard2003,Huenemoerder2003,Sanz-Forcada2003,Argiroffi2004}.  There is a clear relation between the FIP bias (defined by the ratio of elemental abundance in the corona to that in the photosphere) and the spectral type; early G stars all have a solar-like FIP effect, which decreases toward early K stars, reaching no FIP effect at about K5, and then reverses to the i-FIP effect for later than K stars with the magnitude of the effect increasing with spectral type \citep{Wood2010,Wood2018}.  For active stars showing i-FIP effects, the abundances decrease with decreasing FIP.  However, there is emerging evidence that a turnover occurs at a very low FIP below which the elemental abundances steeply increase \citep[e.g.,][]{Huenemoerder2013}.  The origin of the turnover remains unclear.  
f
Substantial efforts have been devoted to interpret the FIP fractionation and there is a consensus that the fractionation results from a separation of ions and neutrals in the upper chromosphere with a temperature of 6000--10000\,K, where low-FIP elements are ionized but high-FIP elements are mostly in neutral states.  Among several theoretical models proposed \citep[e.g.,][]{Henoux1998}, the only model that can explain both FIP and i-FIP effects is the Laming model, which is involved in {\it ponderomotive forces}, a nonlineaer force that a charged particle experiences in an inhomogeneous oscillating electromagnetic field \citep{Laming2004,Laming2009,Laming2012,Laming2015}.  Ponderomotive forces in the chromosphere may in principle be directed upward or downward, resulting in FIP and i-FIP effects, respectively.  Since upward ponderomotive forces are common in solar condition, i.e., the Alfven wave energy flux through the chromosphere is nonzero \citep[see Section 3.2 in][for more details]{Laming2004}, we generally see enhancements of low-FIP elements in the solar corona.  In coronal holes with open-field lines, upward ponderomotive forces are expected to become very weak, and thus a very low FIP effect is expected there, which is indeed observed \citep{Feldman1993}.  In sunspots, where p-mode waves probably convert to fastmode waves and the fastmode waves get reflected back toward the chromosphere \citep{Laming2015}, downward ponderomotive forces are expected.  Therefore, it is reasonable that i-FIP effects are observed in magnetically active stars whose surfaces are largely covered by starspots.

Elemental abundances during solar flares have been reported to be coronal or intermediate between the coronal and photospheric values \citep{Sterling1993,Sylwester1998,Fludra1999,Phillips2012,Narendranath2014,Sylwester2015,Dennis2015}.  However, it was recently realized that the abundance pattern is different from coronal abundances, in that the boundary between low-FIP and high-FIP elements is about 7\,eV \citep{Dennis2015}, significantly less than $\sim$10\,eV that is normally observed for solar coronae or solar energetic particles.  This implies that the fractionation of flare plasmas are cool and expected to be in lower chromospheric altitudes than that of the normal coronal plasma.  In fact, a few observations of solar flares showed nearly photospheric abundances \citep{Veck1981,Warren2014}, indicating that the bulk of the plasma evaporated during a flare comes from deep in the chromosphere \citep{Warren2014}.  

Recently, \citet{Doschek2015} discovered the i-FIP effect on the Sun, for the first time.  They found unexpectedly high Ar XIV / Ca XIV intensity ratios in small regions near sunspots during solar flares.  Since Ar and Ca are high- and low-FIP elements, respectively, the result agrees with the prediction of the Laming model.  One important question is whether the i-FIP effect is caused by an enhancement of the high-FIP element or a depletion of the low-FIP element.  In the Laming model, we expect the latter case, because the ponderomotive force is effective only to low-FIP elements.  \citet{Doschek2016} estimated the path length of the plasma to be smaller than what is expected for the plasma with photospheric abundances, suggesting the depletion of the low-FIP elements.  More recently, \citet{Brooks2018} pointed out that coronal loops showing i-FIP effects persisted for $\sim$40 minutes, which is more than a factor of four longer than the lifetime of the other loops exhibiting normal FIP effects.  This indicates the depletion of the low-FIP elements in the i-FIP regions.

Suzaku was the fifth in a series of Japanese X-ray astronomy satellites, launched on 2005 July 10 \citep{Mitsuda2007}.  It was devoted to observations of celestial X-ray sources, whereas a large amount of Earth albedo of the solar X-ray emission was observed as a by-product, because of its low Earth orbit at 550\,km altitude combined with the 3-D fixed attitude during each observation.  The Earth albedo data were obtained almost every orbit or $\sim$96 minutes, when the telescope pointed at the bright Earth illuminated by the Sun (the duration is typically several minutes every orbit).  Therefore, Suzaku monitored the Sun throughout the 10-year lifetime of its mission with a $\sim$96-minutes cadence.  

Suzaku carried three distinct co-aligned scientific instruments.  The primary instrument was an X-ray micro-calorimeter, which unfortunately became inoperable before performing observations due to unexpected evaporation of liquid helium coolant in the early commissioning phase.  The remaining two instruments, X-ray sensitive imaging CCD cameras \citep[X-ray Imaging Spectrometer, XIS:][]{Koyama2007} and a non-imaging collimated Hard X-ray Detector \citep[HXD:][]{Takahashi2007} worked well through the entire mission.  In this paper, we focus on the thermal X-ray emission from the Sun, concentrating on the data taken by the XIS which covers an energy range of 0.2--12\,keV.

We present absolute elemental abundances of Si, S, Ar, and Ca (relative to H) for four flares with GOES classes from X5.4 to X17.0, based on line-to-continuum ratios.  Absolute abundances are much more difficult to determine than relative abundances, because most of the hydrogen in the corona is ionized and no observable spectral lines are produced.  The amount of hydrogen (and helium) must be obtained by continuum emission.  This is generally difficult with solar high-resolution dispersive spectrometers for several possible reasons, such as instrumental fluorescence background emission, an improper shape for instrumental spectral response \citep{Sylwester2002}.  Some non-dispersive spectrometers had the capability to reliably measure continuum emission, but their spectral resolutions were poor, e.g., FWHM $\sim$ 600\,eV at 5.9\,keV for MESSENGER/SAX \citep{Dennis2015}, FWHM $\sim$ 200\,eV at 5.9\,keV for Chandrayaan-1/XSM \citep{Narendranath2014}, and FWHM $\sim$ 150\,eV in 0.2--4\,keV for Amptek/X123-SDD \citep{Caspi2015}, hampering reliable measurements of absolute abundances.  In this way, significant systematic uncertainties still remain on coronal absolute abundances.  Thanks to the stable and low background of the XIS \citep{Tawa2008} combined with better spectral resolution of $E/\Delta E \sim 20$ or FWHM$\sim$120\,eV at 3\,keV than other non-dispersive spectrometers, we measured the cleanest line-to-continuum ratios and absolute elemental abundances.  We present the data and their reduction in Section 2, and perform spectral analyses in Section 3.  The results are discussed in Section 4.  More details on the data reduction and the simulation to examine spectral deformation due to Earth albedo are given in appendices 1 and 2, respectively.  The errors quoted in this paper are at the 1$\sigma$ confidence level, unless otherwise stated.

\section{Observations and Data Reduction}

\begin{figure}
\figurenum{1}
\plotone{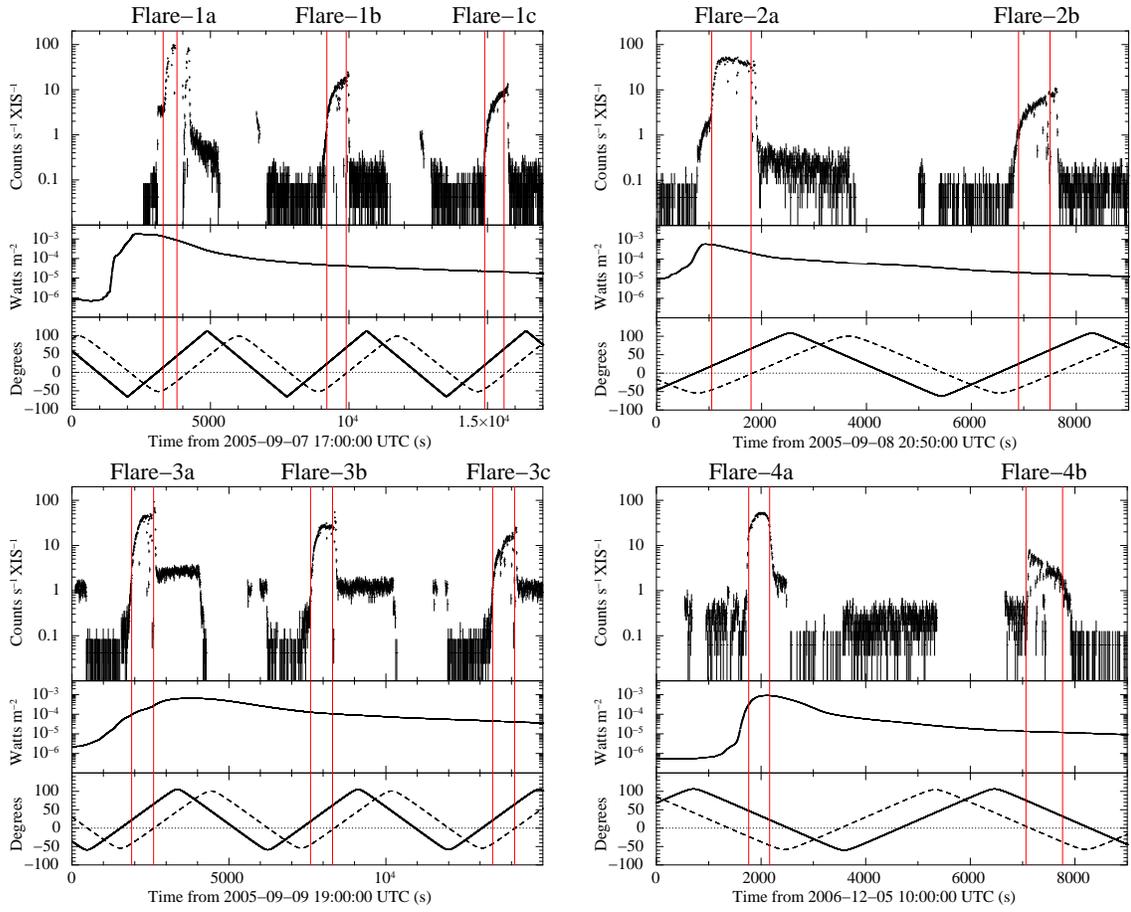} 
\caption{Light curves of the four flares.  Upper panels show XIS data averaged for the two or three front-illuminated CCDs in the 1.65--5\,keV band.  Middle panels show GOES X-ray fluxes in 1--8\,\AA.  Bottom panels show Suzaku/telescope's (dashed lines) and Sun's (solid lines) elevation angles measured from the Earth limb.  The red vertical lines indicate the time periods when we extracted the XIS spectra.}\label{fig:lc}
\end{figure}

Earth albedo emission observed with Suzaku/XIS is spatially unresolved, full disk integrated emission of the Sun, and thus consists of various components such as active regions, quiet regions, and flares.  However, at a large, say GOES X-class, solar flare event, flare emission itself dominates the observed X-ray emission by $\sim$90--99\%, as can be seen from the light curves and spectra in Figures~\ref{fig:lc} and \ref{fig:spec}.  In such cases, we obtain nearly pure flare emission.  In addition, the large X-ray fluxes allow us to obtain enough photons for detailed spectroscopy.  Therefore, we focus on only large flares in this paper.  

Fifty nine X-class flares occurred during the period when Suzaku was in operation.  Among them, 14 flares were captured when Suzaku pointed at the bright Earth.  Although these 14 flares allow us to obtain sufficient photons for our spectral analysis, we further selected early-epoch four flares, i.e., Flares-1, 2, 3, and 4 occurred on 2005 September 7 (X17.0), 2005 September 8 (X5.4), 2005 September 9 (X6.2), and 2006 December 5 (X9.0), respectively.  This is because these early-epoch data have the best spectral resolution; spectral resolution of the XIS degrades with time due to the integrated radiation damage of CCDs.  Quantitatively, we set a criterion that one-sigma width of fluorescent Ar K$\alpha$ is less than 50\,eV.  With this energy resolution, we can clearly separate fluorescent Ar K$\alpha$ and Ar K$\beta$ lines, which is necessary to measure the line intensity of the Ar He$\alpha$ complex located between the two fluorescences.  Throughout this paper, we use only front-illuminated CCDs, i.e., XIS0, 2, 3, because they have slightly better spectral resolution than the back-illuminated CCD, XIS1.  

Information of Suzaku observations used in this paper is summarized in Table~\ref{tab:obs}.  As described in Section~1, the attitude of Suzaku is 3-D fixed, and thus the scattering angle of solar X-rays at the Earth atmosphere is fixed during one observational sequence.  The scattering angle can be calculated by ``180 - (XRT-Sun angle) deg", where the XRT-Sun angle is the angle between the telescope axis and direction to the Sun.  The scattering angles were close with each other, agreeing within 7 degrees for the three observations of our interest.

We reprocessed the XIS data, following the standard screening criteria\footnote{Section 6.3 in http://heasarc.gsfc.nasa.gov/docs/suzaku/analysis/abc} recommended by the calibration team of Suzaku.  One exception is the elevation angle, for which we changed the selection criteria from ``ELV $>$ 5 \& DYE$\_$ELV $>$ 20" (by default) to ``ELV $<$ $-10$ \& NTE$\_$ELV $>$ 5" in order to extract the bright Earth data.  In addition, we developed a new method to infer better backgrounds of the XIS.  We describe detailed techniques of this method in appendix~1.  Briefly, the method re-estimates the dark level more precisely than the standard process does, based on information about 16 pixels surrounding each event island.  This method is also effective to recover spectral resolution, especially for the data taken during time periods showing more rapid time variation of background than the on-board dark estimation, such as in the bright Earth observation.

\begin{figure}
\figurenum{2}
\gridline{\fig{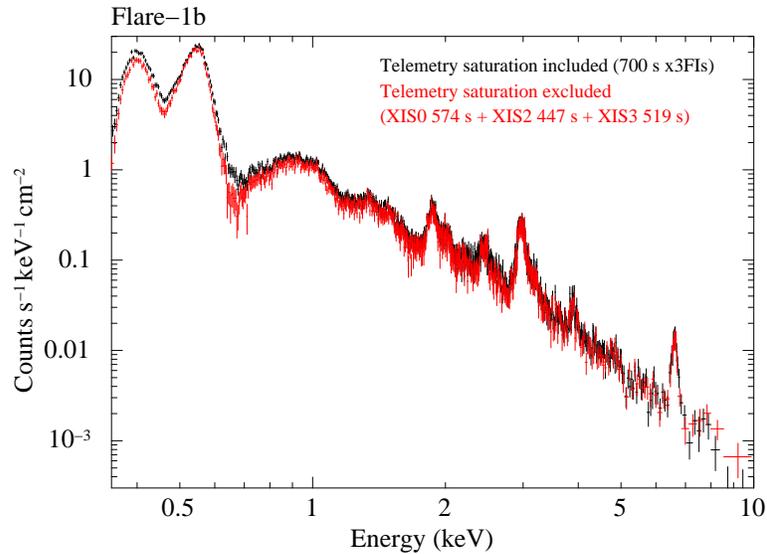}{0.6\textwidth}{}
          }
\caption{XIS spectra (three FIs combined) during Flare-1b.  Black and red correspond to the one before and after excluding the time period of the telemetry saturation.  The spectral quality is essentially the same with each other, except that the effective exposure time decreased by $\sim$30\% after excluding the telemetry-saturation period.}\label{fig:telemetry}
\end{figure}

Figure~\ref{fig:lc}, top panels for each flare section, show XIS light curves in the 1.65--5\,keV band.  They are averaged all over available (two or three) front-illuminated CCDs.  Middle panels show GOES X-ray light curves, and bottom panels show elevation angles of the telescope (dashed lines) and the Sun (solid lines) measured from the Earth's limb.  In the XIS light curves, the high-count-rate periods represent the bright Earth observation, which are confirmed by the elevation angles of the telescope and the Sun.  The long-term variations of XIS light curves in the bright Earth periods follow GOES X-ray light curves.  However, there are significant discrepancies between the XIS and GOES light curves within individual bright Earth periods.  This is because the XIS count rate in the bright Earth period is affected not only by the intrinsic solar X-ray flux but also by the elevation angles of both the telescope and the Sun.  Also, there are abruptly dropping XIS data points in the bright Earth periods.  These are caused by the telemetry saturation.  To check the effect of the telemetry saturation, we compare spectra before and after excluding the telemery-saturation periods for Flare-1b, as shown in Figure~\ref{fig:telemetry}.  This comparison demonstrates that the telemetry saturation does not significantly affect the spectral shape, especially in the energy range of our interest, i.e., 1.65--5\,keV.  Thus, we decided to use telemetry-saturated data in our analysis; otherwise we lost all the exposure times for Flare-1a and Flare-2a.

Our target, bright Earth during giant solar flares, is indeed so bright that one may concern the pile-up effects, i.e., more than one photons strike the same CCD pixel, which generally results in a flux decrease and spectral hardening.  \citet{Maeda2009} investigated the XIS data for the Cassiopeia A supernova remnant, one of the brightest diffuse X-ray sources, and concluded that pile-up does not affect the XIS data.  The XIS count rate of Cassiopeia~A was about 118\,s$^{-1}$ for each front-illuminated CCD.  This is comparable with that for Flare-1a which shows the maximum count rate in the bright Earth.  The emitting area of Cassiopeia~A is about a radius of 3$^\prime$, which is only $\sim$10\% of the bright Earth data extending the entire field of view of the XIS.  Therefore, the surface brightness of the bright Earth is an order of magnitude smaller than that of Cassiopeia~A, leading us to conclude that bright Earth is free from pile-up effects.

As shown in Figure~\ref{fig:lc}, the four flares of interest lasted much longer than Suzaku's orbital period of 96 minutes.  Therefore, we extracted XIS spectra from a few bright Earth periods for each flare (e.g., Flare-1a, Flare-1b, and Flare-1c).  The effective exposure time for each spectrum is given in Table~\ref{tab:obs}.

\section{Spectral Analysis and Results}

\begin{figure}
\figurenum{3}
\gridline{\fig{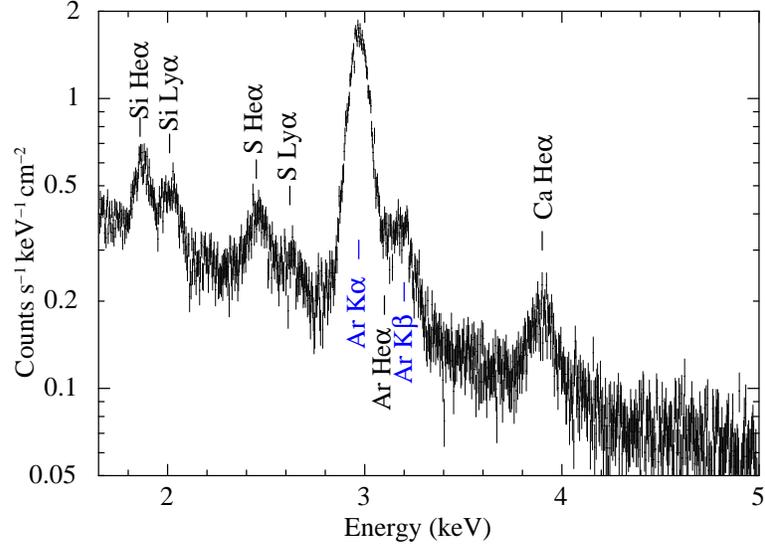}{0.6\textwidth}{}
          }
\caption{XIS spectrum (three FIs combined) during Flare-2a.  This spectrum is before pre-flare subtruction.  Prominent line features are identified.  }\label{fig:spec_demo}
\end{figure}

\begin{figure}
\figurenum{4}
\gridline{\fig{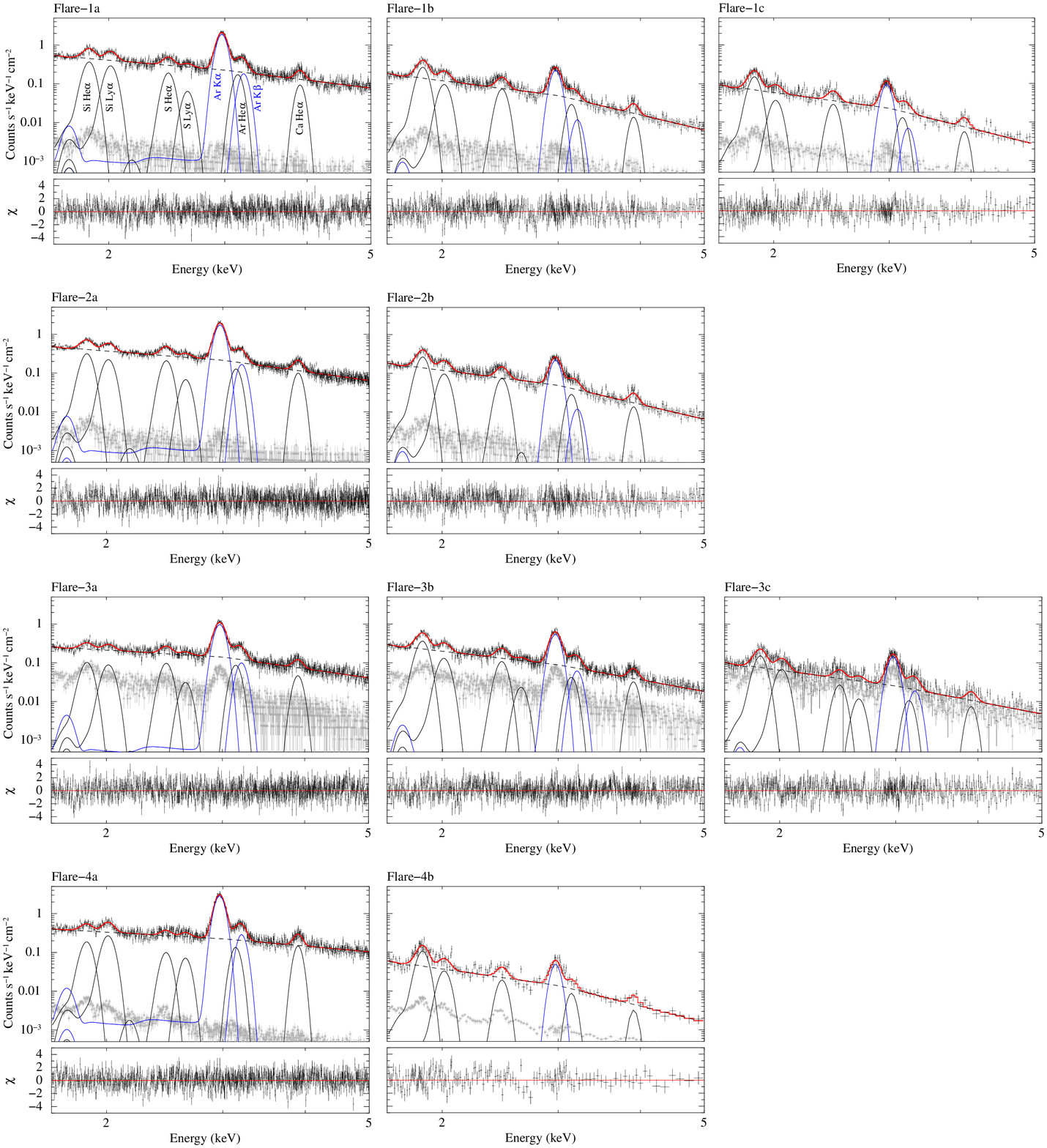}{1\textwidth}{}
          }
\caption{XIS spectra for Flare-1 to 4 from the top to bottom.  Upper panels show the data together with the best-fit model consisting of eight Gaussians plus a broken power-law.  The total emission models are shown in red, whereas thermal lines, continuum, and fluorescent lines are shown in solid black, dashed black, and blue lines, respectively.  The grey data are pre-flare backgrounds.  Bottom panels represent the residuals.}\label{fig:spec}
\end{figure}

For a demonstration, we present in Figure~\ref{fig:spec_demo} Suzaku/XIS's bright Earth spectrum during Flare-2a.  The data are the sum of the three XIS sensors before subtracting the pre-flare background which is only a 1\% level of the source in the energy range of interest (see Figure~\ref{fig:spec}).  There are several prominent features above the underlying continuum emission.  By fitting the spectrum with eight Gaussians plus a broken power-law component, we derived the line centroids, widths, and fluxes summarized in Table~\ref{tab:spec_demo}.  The line centroids allow us to identify them as Si He$\alpha$, Si Ly$\alpha$, S He$\alpha$, S Ly$\alpha$, and Ca He$\alpha$ features together with fluorescent Ar K$\alpha$ and Ar K$\beta$ lines.  The highly-ionized features should originate from the Sun, whereas the fluorescent lines should arise from the Earth atmosphere.  If we expand the energy range of the XIS spectrum, we can identify some other features including fluorescent O K$\alpha$ and N K$\alpha$ around 0.5\,keV as well as Fe He$\alpha$ complex and a possible Compton shoulder hump at $\sim$6.5\,keV.  In this paper, we will focus only on the intermediate-mass elements, i.e., Si, S, Ar, and Ca arising from the solar flares, because lines from these elements are cleaner than those from other elements such as Mg (contaminated by instrumental Al lines) and Fe (contaminated by emission reflected by the satellite).  Analyses of other lines including fluorescent K lines from N, O, and Ar are ongoing, and will be presented elsewhere.

Figure~\ref{fig:spec} presents all the XIS spectra to be analyzed, i.e., the bright Earth data for the four flares.  The data in black and grey are the sums of data and preflare backgrounds, respectively, for which we utilized only front-illuminated CCDs.  The background spectra, shown in grey in Figure~\ref{fig:spec}, are taken from preflare bright Earth for each flare.  Since Flare-1 and Flare-2 occurred continuously, we used the same background taken prior to Flare-1.  Redistribution matrix files (RMF) and auxiliary response files (ARF) were generated by using {\tt xisrmfgen} and {\tt xissimarfgen} \citep{Ishisaki2007}, respectively.  Briefly, the RMF is a probability map from incident photon energy space into detector's calibrated pulse height space, and the ARF gives effective areas as a function of a photon energy.  These two files are convolved with a model spectrum to fit the data.  Details of the RMF and ARF are described in Section~4 in \citet{Ishisaki2007}.

\subsection{Measuring Equivalent Widths}

The spectral shape of solar X-ray emission is expected to be modified by the Earth atmospheric scattering.  This spectral deformation must be considered when interpreting the bright Earth data.  In this work, we focus on equivalent widths (EWs), which are not affected by the spectral deformation, because both lines and their underlying continua are equally deformed.  We measured EWs directly from the original data, and converted them to absolute elemental abundances.  

We note that there is a pioneering work by \citet{Itoh2002} who analyzed the Earth albedo data obtained with ASCA.  To take into account the albedo effects, they adopted the {\tt hrefl} model in XSPEC \citep{Arnaud1996}, which was developed by Dr.\ Yaqoob to consider reflection by neutral matters.  \citet{Itoh2002} assumed it to be a good approximation for the Earth albedo.  This assumption may be good enough.  However, it is technically difficult for the {\tt hrefl} model to take into account the Earth atmospheric abundances.  Therefore, we did not use the {\tt hrefl} model in our analyses, but relied on EWs that are free from spectral deformation.

We fitted the data with a simple combination model consisting of eight Gaussians and a continuum model.  The eight Gaussians include the prominent seven features listed above, and Ar He$\alpha$ that is required to fit the data between fluorescent Ar K$\alpha$ and Ar K$\beta$.  We adopted a broken power-law model for the continuum emission.  In this way, we measured EWs for individual features.  The results are summarized in Table~\ref{tab:ew_obs}.

\subsection{Converting EWs to Abundances}

The EWs were then converted to absolute elemental abundances.  To this end, we calculated model EWs as a function of abundance to compare with our measurements.  We adopted (1) a two-temperature (2$T$) model with two values of $T$ and emission measure ($EM$), and (2) a multi-temperature model with power-law dependence of the differential emission measure (DEM) on the temperature, i.e., $EM$ = constant $\times\ T^{\alpha}$.  Specifically, we used the {\tt vapec} and {\tt cevmkl} models in XSPEC for the 2$T$ and DEM cases, respectively, where both models give an emission spectrum from collisionally-ionized diffuse gas, using the atomic database version of ATOMDB3.0.3.  Ideally, it would be best if we can constrain the two temperatures or the maximum temperature by our data themselves.  This may be possible if we can perform detailed spectral fitting, taking account of the spectral deformation due to Earth albedo effects.  However, such a sophisticated spectral modeling is beyond the scope of this paper.  We thus assumed the temperatures ($kT$) of the 2$T$ model to be 0.5\,keV and 1.7\,keV, based on previous studies showing a bimodal DEM with peak temperatures of those values \citep[e.g.,][]{Sylwester2014,Caspi2014}.  As for the DEM model, we assumed $kT_{\rm max}$ to be 2\,keV, because the lines of our interest (lines from He- and H-like ions of Si, S, Ar, and Ca) are formed by thermal plasmas with a temperature range of 0.2--2\,keV.  For X-class flares, there may be additional super-hot ($kT_{\rm} \gtrsim $ 3\,keV) components \citep[e.g.,][]{Caspi2010}.  This component may contribute to the data to some extent, but not much in the energy range below 5\,keV.  

\begin{figure}
\figurenum{5}
\gridline{\fig{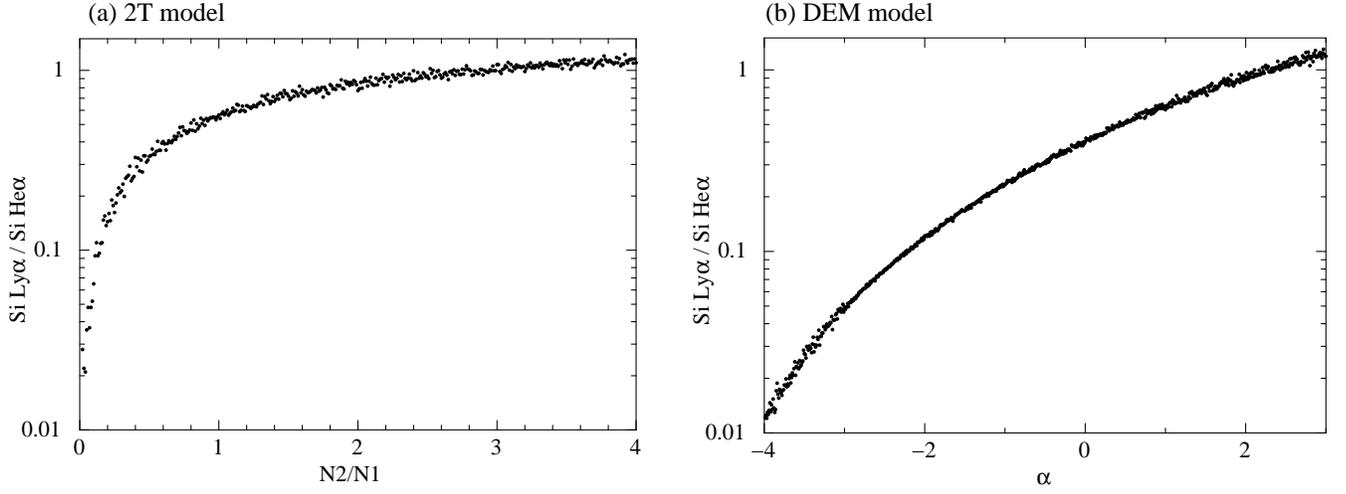}{1\textwidth}{}
          }
\caption{(a) Si Ly$\alpha$ / Si He$\alpha$ ratio for the 2$T$ model.  The x-axis, N2/N1, is a normalization ratio between two {\tt apec} components with temperatures of 1.7\,keV and 0.5\,keV (see text for more details).  The scatter of the data points is due to the statistical fluctuation arising from spectral fits to faked (simulated) XIS spectra.  
(b) Same as left but for the DEM model.  The x-axis, $\alpha$, is the index of the power-law temperature dependence of the EM.}\label{fig:Siratio_vs_N2N1_alpha}
\end{figure}

In our case of the 2$T$ (0.5\,keV + 1.7\,keV) model or the DEM model, the EW depends not only on the elemental abundance but also on either the normalization ratio between the two components (N2/N1) or the index ($\alpha$) of the power-law temperature distribution.  Therefore, to convert EWs to absolute abundances, we need to know these parameters, which are sensitive to, and thus can be measured by intensity ratios between lines from different charge states of the same element.  To this end, we chose a Si Ly$\alpha$ / Si He$\alpha$ intensity ratio, as it is best constrained in our data.  Figure~\ref{fig:Siratio_vs_N2N1_alpha} illustrates how the Si Ly$\alpha$ / Si He$\alpha$ depends on N2/N1 and $\alpha$.  Based on these plots combined with Si Ly$\alpha$ / Si He$\alpha$ ratios, we assessed N2/N1 and $\alpha$ values for all the spectra, as in Table~\ref{tab:Nratio_alpha}.  

It should be noted that the Si Ly$\alpha$ / Si He$\alpha$ ratio increases due to the Earth albedo (the effect of spectral hardening).  To take into account this effect, we performed a Monte Carlo simulation.  Details on the simulation are described in appendix~2.  Briefly, the input of the simulation is the intrinsic solar X-ray spectrum, i.e., the {\tt apec} model with a typical solar flare temperature of $kT = 1.5$\,keV, and the output is an Earth albedo spectrum.  The simulation indicated that the intrinsic Si Ly$\alpha$ / Si He$\alpha$ are 0.8 times the albedo data, namely observations.  We used the corrected Si Ly$\alpha$ / Si He$\alpha$ ratios to infer the values of N2/N1 and $\alpha$ in Table~\ref{tab:Nratio_alpha}.

\begin{figure}
\figurenum{6}
\plotone{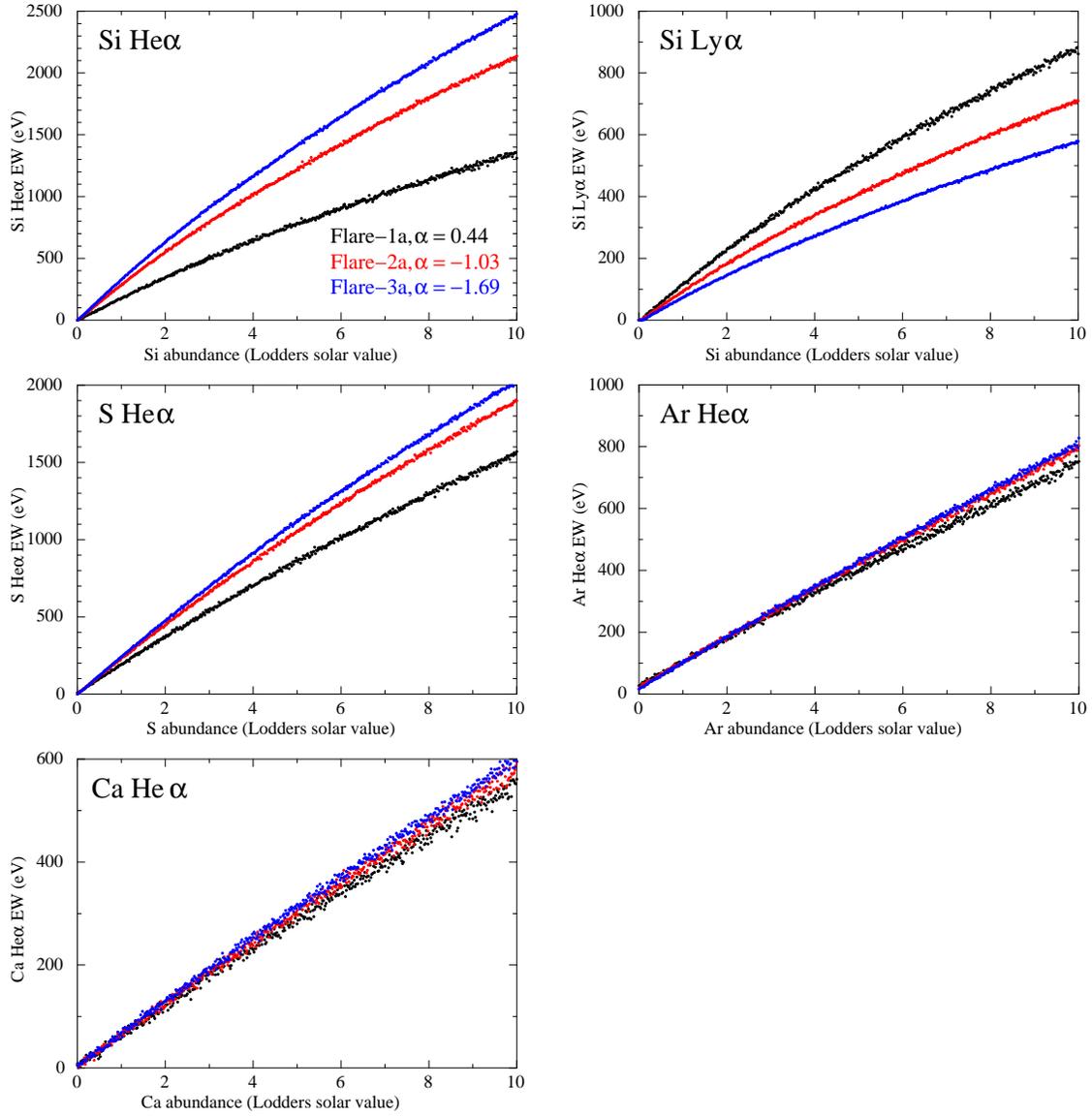} 
\caption{Equivalent widths as a function of metal abundances for Si He$\alpha$, Si Ly$\alpha$, S He$\alpha$, Ar He$\alpha$, and Ca He$\alpha$.  These are based on the DEM model, and the temperature slope parameters ($\alpha$) are set to those for Flare-1 spectra.  Black, red, and blue are responsible for Flare-1a, 1b, and 1c, respectively.  Similarly to Figure~\ref{fig:Siratio_vs_N2N1_alpha}, the scatter of the data points is the statistical fluctuation arising from spectral fits to faked spectra.}\label{fig:model_ew}
\end{figure}

\begin{figure}
\figurenum{7}
\gridline{\fig{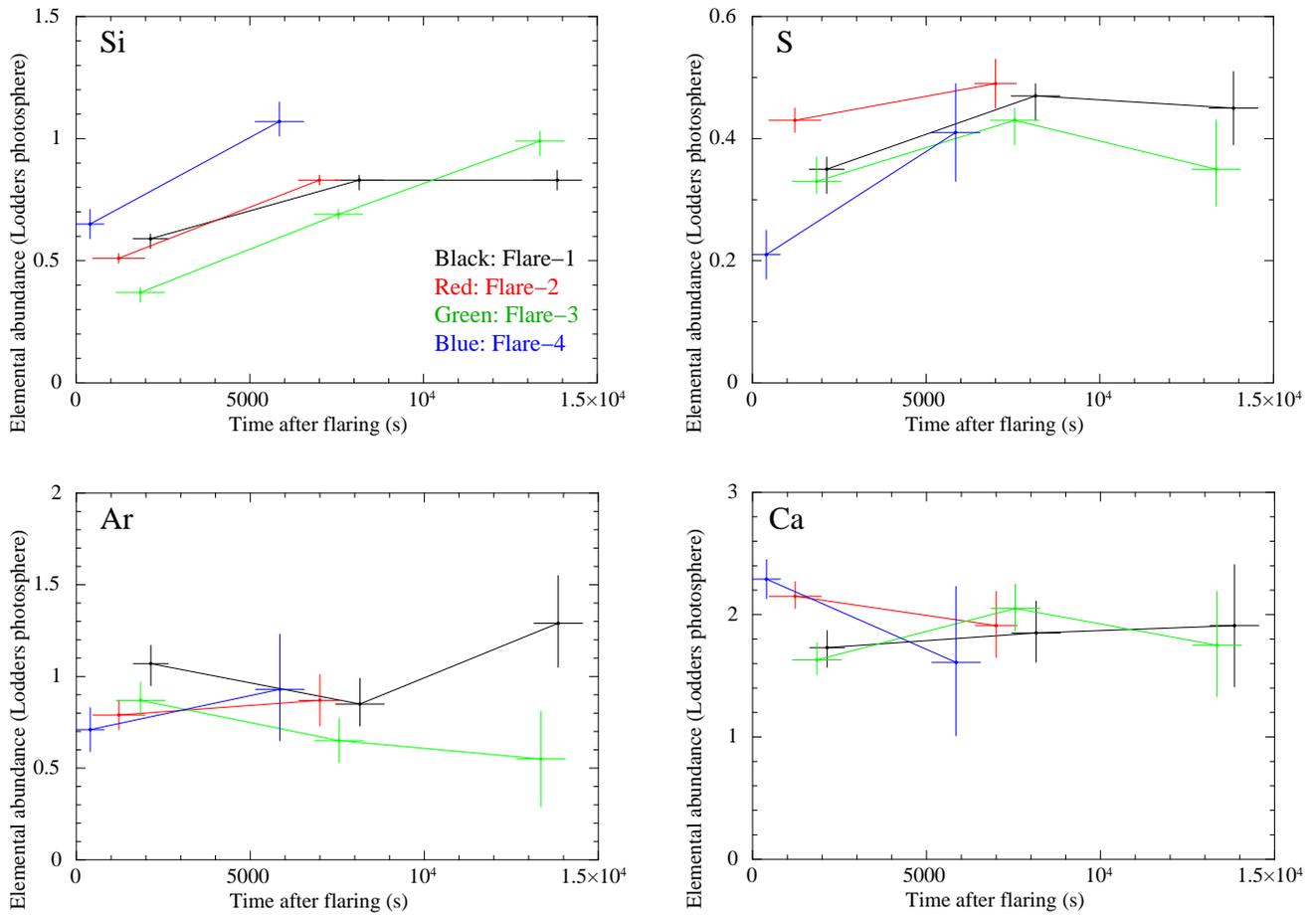}{1.0\textwidth}{}
          }
\caption{Elemental abundance (X/H) relative to the photospheric value as a function of time.  Black, red, green, and blue correspond to Flares-1, 2, 3, and 4, respectively.  These abundances are derived from the DEM model.  The Si abundance is based on the Si He$\alpha$.}\label{fig:abund}
\end{figure}

We computed model EWs as a function of absolute abundances at the best-estimated N2/N1 (2$T$ model) or $\alpha$ (DEM model) for individual spectra from Flare-1a to Flare-4b.  In principle, the model EWs can be derived directly from the emission models, i.e., {\tt vapec} and {\tt cevmkl}.  However, we ``measured" the model EWs by fitting simulated XIS spectra with the same model as we used to measure EWs.  This process is important especially for X-ray CCDs with moderate spectral resolution that cannot resolve individual lines; the Gaussians in our fitting include not only the main line(s) but also some other fainter lines that might affect the EW measurements.  Therefore, we first generated model spectra convolved with the XIS response function, by using the {\tt fakeit} command in XSPEC.  We then fitted the simulated XIS spectra with the same combination model as we did in our EW measurements for real data, obtaining model EWs.  Figure~\ref{fig:model_ew} exhibits DEM-model--based EWs as a function of absolute abundance for Flares-1a, 1b, and 1c.  Using these plots, we converted the EWs to absolute elemental abundances, which are summarized in Table~\ref{tab:abund}.  The uncertainties on the abundances estimated by varying the EWs within their uncertainties.  Throughout this paper, we give the abundances relative to the photospheric values by \citet{Lodders2003}; Si/H = 3.47$\times$10$^{-5}$, S/H = 1.55$\times$10$^{-5}$, Ar/H = 3.55$\times$10$^{-6}$, and Ca/H = 2.19$\times$10$^{-6}$ in number.  These values are in between the classical abundances by \citet{Anders1989} and a more recent abundances by \citet{Asplund2009}; the former gives Si/H = 3.55$\times$10$^{-5}$, S/H = 1.62$\times$10$^{-5}$, Ar/H = 3.63$\times$10$^{-6}$, and Ca/H = 2.29$\times$10$^{-6}$, whereas the latter gives Si/H = 3.24$\times$10$^{-5}$, S/H = 1.32$\times$10$^{-5}$, Ar/H = 2.51$\times$10$^{-6}$, and Ca/H = 2.19$\times$10$^{-6}$.  Figure~\ref{fig:abund} shows the same absolute abundances as a function of time after flaring, where flaring times were defined as GOES X-ray rise-up times determined by eyes.

We confirmed that the Si He$\alpha$- and Si Ly$\alpha$-based Si abundances generally agree with each other.  This shows the robustness of our conversion process from EWs to abundances.  Also, the two abundances derived from the 2$T$ and DEM models are close with each other, showing the stability on the emission model assumed.

\section{Discussion}

\begin{figure}
\figurenum{8}
\gridline{\fig{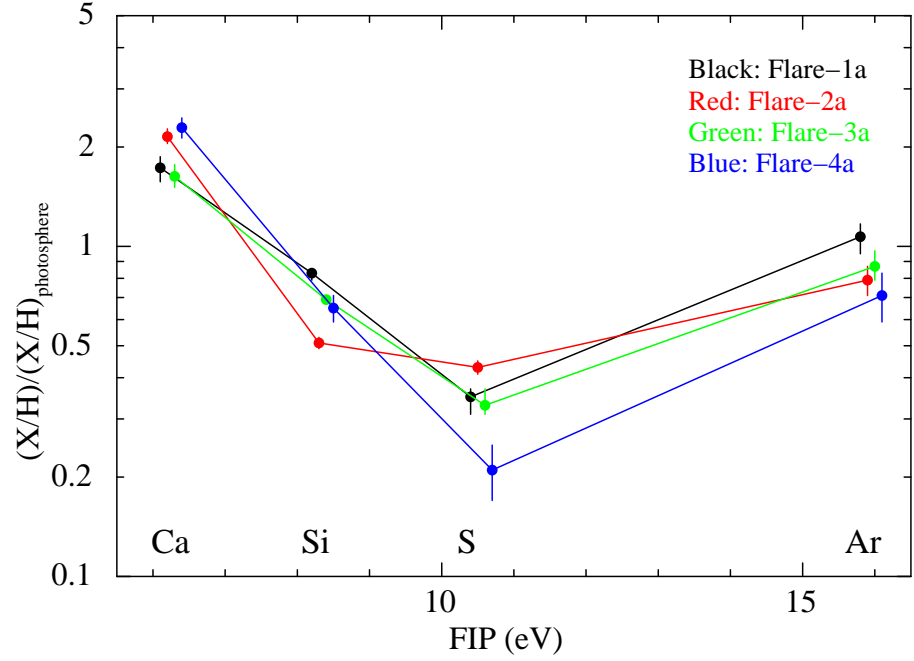}{0.7\textwidth}{}
          }
\caption{Absolute elemental abundances as a function of FIP.  Black, red, green, and blue correspond to Flares-1, 2, 3, and 4, respectively.  For clarity, we introduce an offset in FIP by +0.1\,eV.  These abundances are derived from the DEM model.  The Si abundance is based on the Si He$\alpha$.}\label{fig:abund_vs_fip}
\end{figure}

In Figure~\ref{fig:abund_vs_fip}, we plot the absolute abundances as a function of FIP measured at around each flare peak (i.e., Flare-[1-4]a spectra).  The averaged elemental abundances are calculated to be Ca$\sim$2.0, Si$\sim$0.7, S$\sim$0.3, and Ar$\sim$0.9 solar photospheric values.  This abundance pattern reminds us of the i-FIP effect with a turnover at the low end of FIP as observed in coronae of several active stars \citep{Osten2003,Sanz-Forcada2003,Argiroffi2004,Huenemoerder2013}.  In fact, such i-FIP effects from solar flares are consistent with the picture that high/low-FIP enriched coronae feature a relative enrichment of low/high-FIP elements during flares \citep{Nordon2008}.  We note that their suggestion is mainly based on Chandra and XMM-Newton observations of stars, while they did include the Sun as one of the examples of low-FIP enriched coronae and high-FIP enhanced (relative to low-FIP elements) flares.  

In the decay phase of the flares, the Si abundance significantly increases up to $\sim$2 times the initial value, so that the abundance approaches the coronal abundance.  The same trend can be seen for S as well.  This abundance evolution is also seen for flares in other stars; the abundance pattern exhibits the strong i-FIP/FIP effect during flare peaks, and then evolves toward the preflare FIP/i-FIP basal state \citep[e.g.,][]{Tsuru1989,Stern1992,Pan1997,Tsuboi1998,Favata2000,Liefke2010}.  These properties can be explained by the scenario that the excess emission of flares is caused by the chromospheric evaporation, as discussed in \citet{Nordon2008}.

A similar Fe (a low-FIP element with FIP = 7.9\,eV) abundance increase during solar flares were found by \citet{Warren2014} who measured absolute abundances for 21 M9.3 to X6.9 class flares using SDO/EVE spectra.  On the other hand, such low-FIP increases were not seen in many other observations of solar flares \citep[e.g.,][]{Narendranath2014,Dennis2015,Sylwester2015}.  The reason of such flare-to-flare variations is unclear, but one notable difference is the flare size; those investigated by this work and \citep{Warren2014} are generally larger than those by others.  We speculate that the composition of larger flares tends to show a greater departure (or a clearer i-FIP effects) compared with those in coronae.  As a result, they may show significant abundance evolution toward preflare states in the decay phase.

It is interesting to compare our results with i-FIP patches recently discovered near sunspots during solar flares \citep[e.g.,][]{Doschek2015}.  First, the composition is different between the i-FIP patches and the flares of our interest.  The i-FIP patches exhibit anomalously high Ar XIV / Ca XIV intensity ratios, with an extreme case that the measured relative Ar/Ca abundance is about 10 (30) times the photospheric (coronal) value.  This does not agree with our measured Ar/Ca ratios of about 3 times the solar photospheric value.  In addition, the absolute abundance of Ca (Ca/H) appears to differ, as well.  At the i-FIP patches, there are two independent arguments that the Ca abundance is depleted, which makes the relative Ar/Ca abundance very high, as observed \citep{Doschek2016,Brooks2018}.  This would disagree with our measurement, i.e., Ca/H $\sim$ 2 times the photospheric value.  These composition differences are not surprising, given that our data are spatially unresolved, full disk integration, whereas the i-FIP patches are highly-localized small regions.  Perhaps, the i-FIP patches would not make a significant contribution to the total X-ray emission from flares.

The composition difference between i-FIP patches and our measurements could be explained by different turnover FIPs, by analogy to various turnover FIPs observed in active stars: a turnover above Al (FIP = 6.0\,eV), Ca (FIP = 6.1\,eV), Na (FIP = 5.1\,eV), and K (FIP = 4.3\,eV) for $\sigma^2$ CrB \citep{Osten2003}, AB Dor \citep{Sanz-Forcada2003}, PZ Tel \citep{Argiroffi2004}, and $\sigma$ Gem \citep{Huenemoerder2013}, respectively.  Namely, the i-FIP patches would have a particularly low turnover FIP below Ca, whereas most of the flaring region (responsible for our observations) would have a relatively high turnover FIP between Ca and Si (FIP = 8.2\,eV).  Understanding the origin of such a turnover and its spatial variation is left as an important future work.  

One important question about the i-FIP patches in past observations was which element is enhanced and which element is depleted.  Based on direct measurements of EWs, combined with emission models of either a two-temperature model or a multi-temperature model with power-law dependence of the differential emission measure on the temperature, we revealed that the Si abundance is depleted around flare peaks when the i-FIP effect becomes evident.  This is consistent with the ponderomotive fractionation model proposed by \citet{Laming2004,Laming2012,Laming2015,Laming2017}, in which the ponderomotive force, which can be directed either upward or downward depending on the situation, can act only on low-FIP elements, so that enhanced and depleted low-FIP elements are expected for FIP and i-FIP effects, respectively, with high-FIP elements unmodified.  The transition from depletion to enhancement of the Si abundance in the decay phases of the flares would be caused by plasma mixing with the surrounding FIP biased plasmas, as was proposed to explain fading i-FIP patches \citep{Baker2019}.

We point out that the absolute abundances vary from flare to flare by a factor of 2 with no correlations with other physical parameters such as the flare class and temperature.  This is consistent with previous results \citep[e.g.,][]{Sylwester1998}.  On the other hand, we argued that i-FIP effects may be present only for giant flares in their early-phases.  Therefore, there might be an anti-correlation between the absolute abundance and the flare class, if we accumulate a large range of flare classes and focus on the initial phase of flares.  

Finally, we note that the X-Ray Imaging Spectroscopy Mission \citep[XRISM:][]{Tashiro2018}, the Japan-US X-ray astronomy mission scheduled to be launched in early 2022, will dramatically improve abundance studies for solar flares and corona.  XRISM will be in the same low-Earth orbit as Suzaku, allowing for a long-term, 96-minutes cadence monitoring of the Sun via Earth albedo as a byproduct.  Thanks to the superior energy resolution, FWHM$\lesssim$7\,eV (or $E/\Delta E \sim 600$ at 3\,keV) with little energy dependence in 0.2--12\,keV, of the X-ray micro-calorimeter onboard XRISM, we will be able to detect much more lines than Suzaku/XIS.  These include lines from many elements such as C, N, O, Ne, Na, Mg, Al, Fe, and Ni in addition to the four abundant intermediate-mass elements analyzed in this paper.  Therefore, we will be able to obtain detailed abundance patterns from individual flares, as with the cases of the high-resolution X-ray spectroscopy of stellar coronae with gratings onboard Chandra and XMM-Newton.  Furthermore, fluorescent lines of N, O, Ar from the Earth atmosphere will be clearly detected.  These will provide us with intriguing information about upper atmosphere of the Earth.

\section{Conclusion}

We performed X-ray spectroscopy of four X-class flares, using Earth albedo data obtained with Suzaku/XIS.  The good energy resolution and low background of Suzaku/XIS allow several features and underlying continuum emission to be clearly detected.  We measured EWs of line features from intermediate-mass elements, i.e., Si, S, Ar, and Ca, and converted them to the absolute abundances (X/H).  The absolute elemental abundances averaged over the four flares are obtained to be $\sim$2.0 (Ca), $\sim$0.7 (Si), $\sim$0.3 (S), and $\sim$0.9 (Ar) at around flare peaks.  This abundance pattern is consistent with i-FIP effects seen in active stars' coronae.  The depletion of Si (and S) is consistent with the Laming model to explain both FIP and i-FIP effects observed in stellar coronae.  The abundances are almost constant during the flares, whereas Si increases by a factor of $\sim$2 in the decay phase.  Such Si abundance evolution is consistent with a picture that the i-FIP plasma originates from the chromosphere evaporation and then mixes with the surrounding low-FIP biased materials.

\begin{acknowledgments}
We thank Drs.\ Hideki Uchiyama and Makoto Sawada for discussions about self-charge-filling effects of the XIS and the equivalent widths using XSPEC and SPEX codes, respectively.  This work was supported by the Japan Society for the Promotion of Science KAKENHI grant numbers JP17H02864 (SK), 16H03983 (KM), 17K05393 (KS), 18K18767, 19H01908, 19H00696 (KH), 18H01256 (HN).  This work was partly supported by Leading Initiative for Excellent Young Researchers, MEXT, Japan.
\end{acknowledgments}

\appendix 
\section{A new method to estimate the dark level of the XIS}

The telemetry of each event in the Suzaku XIS data analyzed here contains the dark-subtracted pulse-height (PH) values of 9~pixels, which are the event center pixel and the neighboring 8 pixels. The energy value of each event is reconstructed from the 9 PH values in the ground processing based on the grade method\footnote{See the Suzaku technical description for the detail, https://heasarc.gsfc.nasa.gov/docs/suzaku/prop\_tools/suzaku\_td/}. The event telemetry also contains two kinds of information about the surrounding 16 pixels, \textit{pOuterMost} and \textit{sumOuterMost}. \textit{pOuterMost} is a 16-bit ``hit pattern'' identifying which surrounding pixels have PH value grater than the threshold.  \textit{pOuterMost} is also used for event grading. \textit{sumOuterMost} is the summed PH value of all the surrounding pixels that have PH value less than or equal to the threshold. Note that the value of \textit{sumOuterMost} would be around zero within statistical fluctuation if the on-board dark estimation follows the actual CCD pedestal value reasonably well. If not, \textit{sumOuterMost} would have significant offset from zero either positively or negatively. This situation could happen in the case that the background level varies more rapidly than the on-board dark estimation could follow.  Since the number of surrounding pixels contributing \textit{sumOuterMost} can be known from \textit{pOuterMost}, we can measure the amount of the deviation of the on-board dark estimation from the actual CCD pedestal value per pixel, correct the PH values of the 9~central pixels by the amount, and recalculate the energy value of the event based on the corrected PH values.

\begin{figure}
\figurenum{A1}
\gridline{\fig{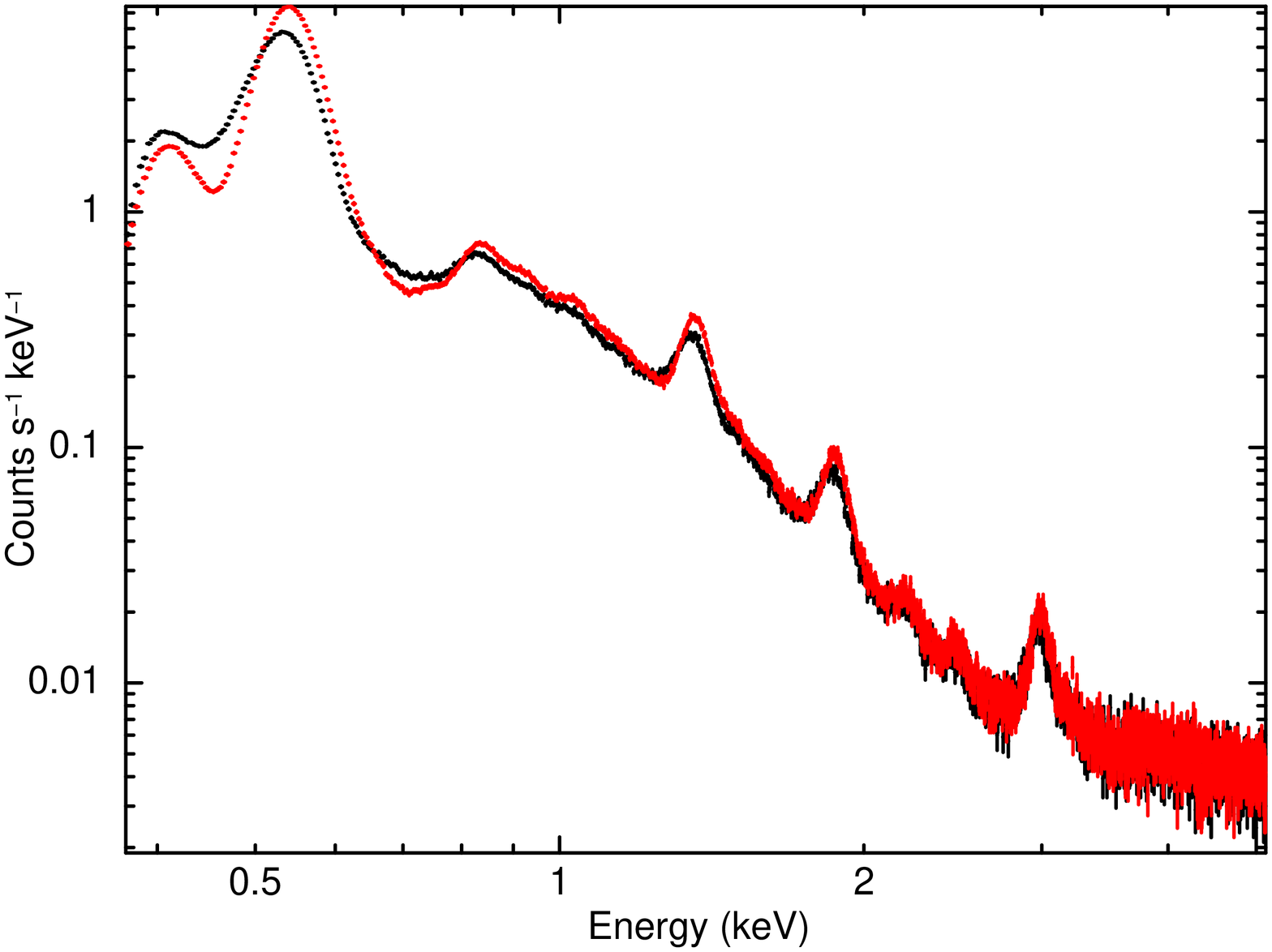}{0.7\textwidth}{}
          }
 \caption{Comparison of the bright Earth data spectra before (black) and after (red) correction. This is the XIS0 case. }\label{fig:ap1}
\end{figure}

Figure~\ref{fig:ap1} shows a comparison of the bright Earth data spectra before and after applying the correction described here.  All the line structures become sharper after the correction, indicating the validity of this method. The sharpness is more evident in the lower energy lines since the ratio of the amount of the correction to the event PH is relatively larger in the lower energy events. We confirm that this method is efficient only for the bright Earth data; no improvement can be seen for the standard data in which the satellite is pointing to the astronomical objects or the dark side of the Earth. This fact indicates that the on-board dark estimation of the Suzaku XIS works well for the standard data and that the intense optical light flux, which is the major source to change the CCD pedestal value in the bright Earth data, varies rapidly than the on-board dark estimation.

\section{Monte Carlo simulation to take account of Earth albedo effects}

The spectral hardening due to the reflection by the Earth albedo would expect to increase the Si Ly$\alpha$ / Si He$\alpha$ line ratio and this effect should be taken into account when interpreting the data.  Since the calculation procedures of the reflected X-ray spectrum by the Earth albedo is complicated due to various scattering processes such as Thomson scattering and Compton scattering, a Monte Carlo simulation including the intrinsic solar X-ray spectrum and a relevant air composition of the Earth is an appropriate approach to estimate an expected scattered X-ray spectrum. In this paper, we performed a Monte Carlo simulation based on the GEANT4 toolkit library (version 10.04) \citep{Apostolakis2003,Allison2006,Allison2013}, which is commonly used to perform a full Monte Carlo simulations including complicated geometries.  In this simulation framework, we input photons with the energy spectrum from collisionally-ionized diffuse gas, named as {\tt apec}, which is calculated by the atomic database with the XSPEC software package \citep{Arnaud1996}.  The atomic database version for the simulation is ATOMDB2.0.2, which is different from that used for the X-ray data analysis in \S 3.1.  We confirmed that the model spectrum in 1--5\,keV shows no significant differences between ATOMDB2.0.2 and 3.0.3, and thus we conclude that our simulation using ATOMDB2.0.2 can be applied for comparison to the data analysis. 

We assume a typical solar flare temperature of $kT =$1.5 keV (see \S~3.1) with an abundance of solar photospheric values.  Photons are generated enough far from the Earth like a point source.  The generated photon interacts the Earth atmosphere in the simulation.  The Earth's atmosphere is composed of oxygen and nitrogen atoms and molecules, as well as argon atoms, which were all included in the simulation.  The density profile of each component is taken into account by implementing the 10\,km thickness of sphere-shell-shape geometry up to 300\,km altitude.  It is known that the density profile would change slightly depending on the solar activity.  We confirmed that our simulation result is not affected by this effect.  Therefore, we apply the density profile obtained during a solar maximum period, August 1st, 2013, which is provided by the publicly available NRLMSISE-00 atmosphere model.\footnote{https://ccmc.gsfc.nasa.gov/modelweb/models/nrlmsise00.php \citep{Labitzke1985,Hedin1991,Picone2002}}

\begin{figure}[htbp]
\figurenum{B1}
\plotone{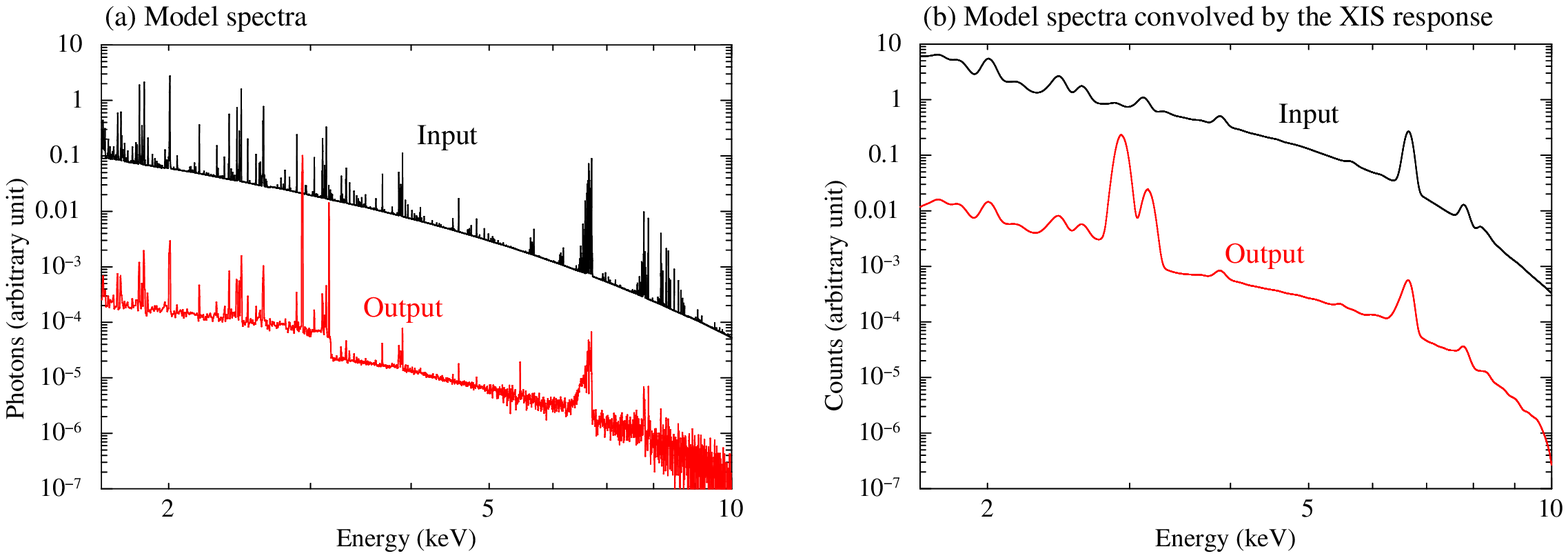} 
\caption{(a) Comparison of the photon spectrum emitted from ionized diffuse gas with a temperature of 1.5 keV calculated by the {\tt apec} model without scattering effect (black) to the simulated photon spectrum after scattering by the Earth atmosphere (red).  (b) Same as left, but the data are convolved by the detector response of Suzaku XIS.}
\label{fig:ap2}
\end{figure}

Since what we have to know is the reflected X-ray spectrum by the Earth albedo, we accumulated photons only scattered by the Earth atmosphere. For this purpose, we put the photon detector at 500\,km altitude in the simulation and record the photon only with the angle between incident and detected position of larger than $\pi$/2.  Figure~\ref{fig:ap2} (a) shows the example of the input and the scattered X-ray spectrum, in which we can see that the hardening of the continuum as well as some changes of line ratios due to the scattering.  The simulated X-ray reflection spectrum is converted to the XSPEC table model, so that we can estimate the line ratio by convolving with the detector response of the Suzaku XIS as shown in Figure~\ref{fig:ap2} (b).  It should be noted that the solar X-ray scattering angles are around 110\,deg in our cases (see Table~1).  We confirmed, however, that there is no obvious spectral difference around 2\,keV between reflection angles of 90--180\,deg and 105--115\,deg. 


\vspace{-0.3cm}
\begin{deluxetable}{cccccc}
 \tablecaption{Suzaku observations used in this paper \label{tab:obs}}
\tablenum{1}
 \tablehead{
     \colhead{Obs.\ ID} & \colhead{Flare ID} & \colhead{Start time\tablenotemark{a}} & \colhead{XRT-Sun angle (deg)} & \colhead{Scattering angle (deg)} & \colhead{Effective exposure time (s)\tablenotemark{b}} 
}
\startdata
      100019010 & Flare-1a & 2005-09-07 17:54:50 & 75.1 & 104.9 & 315 ($\times$3 CCDs) \\
      100019010 & Flare-1b & 2005-09-07 19:33:20 & 75.1 & 104.9 & 700 ($\times$3 CCDs) \\
      100019010 & Flare-1c & 2005-09-07 21:08:20 & 75.1 & 104.9 & 700 ($\times$3 CCDs) \\
      100019010 & Flare-2a & 2005-09-08 21:07:30 & 75.1 & 104.9 & 750 ($\times$3 CCDs) \\
      100019010 & Flare-2b & 2005-09-08 22:45:00 & 75.1 & 104.9 & 700 ($\times$3 CCDs) \\
      100019020 & Flare-3a & 2005-09-09 19:31:40 & 70.9 & 109.1 & 700 ($\times$3 CCDs) \\
      100019020 & Flare-3b & 2005-09-09 21:06:40 & 70.9 & 109.1 & 700 ($\times$3 CCDs) \\
      100019020 & Flare-3c & 2005-09-09 22:43:20 & 70.9 & 109.1 & 700 ($\times$3 CCDs) \\
      801064010 & Flare-4a & 2006-12-05 10:29:24 & 68.1 & 111.9 & 400 ($\times$2 CCDs) \\
      801064010 & Flare-4b & 2006-12-05 11:57:44 & 68.1 & 111.9 & 700 ($\times$2 CCDs) 
\enddata
\tablenotetext{a}{Starting time when we extract the spectrum.} 
\tablenotetext{b}{The entire imaging area of XIS2 was lost in 2005 November possibly due to micro-meteorite hits, so that only two XIS sensors were in operation during Flare-4.}
\end{deluxetable}

\vspace{-0.5cm}
\begin{deluxetable}{ccccccc}
\tablecaption{Line properties measured for Flare-2a \label{tab:spec_demo}}
\tablenum{2}
\tablehead{
     \colhead{Line ID} &  \colhead{Ion} &  \colhead{Expected energy (keV)$^a$} & \colhead{Centroid (keV)} & \colhead{Broadening (keV)} & \colhead{Intensity (10$^{-5}$ ph\,cm$^{-2}$\,s$^{-1}$\,arcmin$^{-2}$)} 
}
\startdata
      Si He$\alpha$ & Si$^{12+}$ & 1.839 (f), 1.854+1.855 (i), 1.865 (r) & 1.869$\pm$0.003 & 0.016$\pm$0.007 & 2.63$^{+0.20}_{-0.18}$ \\
      Si Ly$\alpha$ & Si$^{13+}$ & 2.006 & 2.014$\pm$0.005 & 0.014$\pm$0.009 & 1.48$^{+0.15}_{-0.14}$ \\
      S He$\alpha$ & S$^{13+}$ & 2.430 (f), 2.447+2.449 (i), 2.461 (r) & 2.460$\pm$0.006 & 0.028$\pm$0.011 & 1.672$^{+0.21}_{-0.19}$  \\
      S Ly$\alpha$ & S$^{14+}$ & 2.623 & 2.628$^{+0.014}_{-0.017}$ & 0.019 $(<0.048)$ & 0.54$^{+0.19}_{-0.15}$ \\
      Ar He$\alpha$ & Ar$^{14+}$ & 3.104 (f), 3.124+3.126 (i), 3.140 (r)  & 3.135$^{+0.034}_{-0.021}$ & = Ca~He$\alpha$ & 1.28$^{+0.86}_{-0.22}$\\
      Ca He$\alpha$ & Ca$^{15+}$ & 3.861 (f), 3.883+3.888 (i), 3.902 (r) & 3.902$\pm$0.006 & 0.035$\pm$0.011 & 1.03$^{+0.10}_{-0.10}$\\ 
      Ar K$\alpha$ & Ar$^{0+}$ & 2.958 & 2.972$\pm$0.001 & 0.021$\pm$0.002 & 14.70$^{+0.28}_{-0.27}$ \\
      Ar K$\beta$ & Ar$^{0+}$ & 3.191 & =Ar K$\alpha$ + 0.235 & = Ar K$\alpha$ & 1.55$^{+0.27}_{-0.92}$ 
\enddata
\tablenotetext{a}{(f), (i), and (r) represent forbidden, intercombination, and resonance lines, respectively.} 
\end{deluxetable}

\vspace{-0.5cm}
\begin{deluxetable}{cccccc}
\tablecaption{Equivalent widths from Gaussians plus a continuum model\tablenotemark{a} \label{tab:ew_obs}}
\tablenum{3}
\tablehead{
     \colhead{Flare ID} & \colhead{Si He$\alpha$} & \colhead{Si Ly$\alpha$} & \colhead{S He$\alpha$} & \colhead{Ar He$\alpha$} & \colhead{Ca He$\alpha$} 
}
\startdata
      Flare-1a & 103$\pm$5 & 72$\pm$4 & 71$\pm$5 & 109$\pm$7 &105$\pm$8  \\
      Flare-1b & 237$\pm$7 & 80$\pm$5 & 108$\pm$7 & 92$\pm$10 & 117$\pm$15 \\
      Flare-1c & 276$\pm$13 & 62$\pm$9 & 107$\pm$14 & 125$\pm$20 & 123$\pm$31 \\
      Flare-2a & 96$\pm$3 & 62$\pm$2 & 86$\pm$3 & 87$\pm$5 & 131$\pm$5 \\
      Flare-2b & 234$\pm$7 & 83$\pm$5 & 112$\pm$7 & 92$\pm$10 & 119$\pm$15 \\
      Flare-3a & 55$\pm$5 & 44$\pm$4 & 67$\pm$5 & 94$\pm$7 & 98$\pm$7 \\
      Flare-3b & 195$\pm$6 & 69$\pm$4 & 97$\pm$6 & 75$\pm$9 & 129$\pm$11 \\
      Flare-3c & 243$\pm$12 & 112$\pm$9 & 80$\pm$14 & 67$\pm$21 & 109$\pm$25 \\
      Flare-4a & 74$\pm$6 & 81$\pm$5 & 40$\pm$5 & 80$\pm$8 & 135$\pm$8 \\
      Flare-4b & 334$\pm$20 & 90$\pm$12 & 95$\pm$18 & 97$\pm$24 & 103$\pm$37
\enddata
\tablenotetext{a}{The values are in units of eV.}
\end{deluxetable}

\vspace{-1cm}
\begin{deluxetable}{ccccc}
\tablecaption{Si Ly$\alpha$ / Si He$\alpha$ ratios, N2/N1 for the 2$T$ model, and $\alpha$ for the DEM model \label{tab:Nratio_alpha}}
\tablenum{4}
\tablehead{
      \colhead{Flare ID} & \colhead{Si Ly$\alpha$ / Si He$\alpha$} & \colhead{Albedo-corrected Si Ly$\alpha$ / Si He$\alpha$}  & \colhead{N2 / N1} & \colhead{$\alpha$} 
}
\startdata
      Flare-1a & 0.63$\pm$0.05 & 0.50$\pm$0.04 & 0.84$^{+0.10}_{-0.09}$ & 0.44$\pm$0.17 \\
      Flare-1b & 0.29$\pm$0.02 & 0.23$\pm$0.02 & 0.32$^{+0.02}_{-0.04}$ & -1.03$^{+0.11}_{-0.12}$ \\
      Flare-1c & 0.19$\pm$0.03 & 0.15$\pm$0.02 & 0.20$^{+0.03}_{-0.04}$ & -1.69$^{+0.22}_{-0.23}$ \\
      Flare-2a & 0.58$\pm$0.03 & 0.46$\pm$0.03 & 0.75$^{+0.06}_{-0.05}$ & 0.27$\pm$0.12 \\
      Flare-2b & 0.30$\pm$0.02 & 0.24$\pm$0.02 & 0.33$^{+0.03}_{-0.02}$ & -0.96$\pm$0.12 \\
      Flare-3a & 0.73$\pm$0.09 & 0.59$\pm$0.08 & 1.06$^{+0.23}_{-0.19}$ & 0.80$^{+0.30}_{-0.31}$ \\
      Flare-3b & 0.30$\pm$0.02 & 0.24$\pm$0.02 & 0.34$\pm$0.03 & -0.95$\pm$0.12 \\
      Flare-3c & 0.38$\pm$0.04 & 0.31$\pm$0.03 & 0.45$\pm$0.06 & -0.53$^{+0.18}_{-0.19}$ \\
      Flare-4a & 1.01$\pm$0.11 & 0.80$\pm$0.09 & 1.82$^{+0.43}_{-0.36}$ & 1.62$^{+0.31}_{-0.33}$ \\
      Flare-4b & 0.22$\pm$0.03 & 0.18$\pm$0.03 & 0.24$\pm$0.04 & -1.44$^{+0.22}_{-0.24}$ 
\enddata
\end{deluxetable}

\vspace{-0.5cm}
\begin{deluxetable}{cccccc}
\tablecaption{Elemental abundances\tablenotemark{a} \label{tab:abund}}
\tablenum{5}
 \tablehead{
     \colhead{Flare ID (emission model)} & \colhead{Si He$\alpha$} & \colhead{Si Ly$\alpha$} & \colhead{S He$\alpha$} & \colhead{Ar He$\alpha$} & \colhead{Ca He$\alpha$} 
}
\startdata
      Flare-1a (2T) & 0.77$^{+0.03}_{-0.04}$ & 0.81$\pm$0.04 & 0.47$^{+0.03}_{-0.04}$ & 1.06$\pm$0.11 & 1.48$^{+0.12}_{-0.14}$ \\
      Flare-1a (DEM) & 0.59$^{+0.02}_{-0.04}$ & 0.63$\pm$0.04 & 0.35$^{+0.02}_{-0.04}$ & 1.07$^{+0.10}_{-0.12}$ & 1.73$^{+0.14}_{-0.16}$ \\
      Flare-1b (2T) & 1.06$\pm$0.04 & 1.15$\pm$0.07 & 0.68$\pm$0.05 & 0.92$\pm$0.14 & 1.69$^{+0.25}_{-0.24}$ \\
      Flare-1b (DEM) & 0.83$^{+0.02}_{-0.04}$ & 0.87$^{+0.06}_{-0.04}$ & 0.47$^{+0.02}_{-0.04}$ & 0.85$^{+0.14}_{-0.12}$ & 1.85$^{+0.26}_{-0.24}$ \\
      Flare-1c (2T) & 0.99$\pm$0.05 & 1.06$\pm$0.15 & 0.62$\pm$0.08 & 1.48$\pm$0.29 & 1.83$^{+0.52}_{-0.51}$ \\
      Flare-1c (DEM) & 0.83$\pm$0.04 & 0.87$^{+0.12}_{-0.14}$ & 0.45$\pm$0.06 & 1.29$^{+0.26}_{-0.24}$ & 1.91$\pm$0.50 \\
      Flare-2a (2T) & 0.68$^{+0.02}_{-0.03}$ & 0.71$^{+0.03}_{-0.02}$ & 0.57$\pm$0.03 & 0.77$^{+0.08}_{-0.07}$ & 1.88$^{+0.09}_{-0.10}$ \\
      Flare-2a (DEM) & 0.51$\pm$0.02 & 0.55$\pm$0.02 & 0.43$\pm$0.02 & 0.79$\pm$0.08 & 2.15$^{+0.12}_{-0.10}$ \\
      Flare-2b (2T) & 1.06$\pm$0.03 & 1.17$^{+0.07}_{-0.08}$ & 0.69$^{+0.05}_{-0.04}$ & 0.95$\pm$0.15 & 1.69$^{+0.27}_{-0.25}$ \\
      Flare-2b (DEM) & 0.83$\pm$0.02 & 0.89$^{+0.04}_{-0.06}$ & 0.49$\pm$0.04 &  0.87$\pm$0.14 & 1.91$^{+0.28}_{-0.26}$ \\
      Flare-3a (2T) & 0.48$\pm$0.04 & 0.49$^{+0.04}_{-0.03}$ & 0.45$\pm$0.04 & 0.82$^{+0.10}_{-0.09}$ & 1.33$^{+0.13}_{-0.11}$ \\
      Flare-3a (DEM) & 0.37$^{+0.02}_{-0.04}$ & 0.39$\pm$0.04 & 0.33$^{+0.04}_{-0.02}$ & 0.87$^{+0.10}_{-0.08}$ & 1.63$^{+0.14}_{-0.12}$ \\
      Flare-3b (2T) & 0.89$^{+0.02}_{-0.03}$ & 0.97$\pm$0.06 & 0.61$^{+0.03}_{-0.05}$ & 0.70$^{+0.14}_{-0.13}$ & 1.84$^{+0.20}_{-0.18}$ \\
      Flare-3b (DEM) & 0.69$\pm$0.02 & 0.73$^{+0.06}_{-0.04}$ & 0.43$^{+0.02}_{-0.04}$ & 0.65$\pm$0.12 & 2.05$^{+0.20}_{-0.18}$ \\
      Flare-3c (2T) & 1.27$^{+0.07}_{-0.06}$ & 1.42$\pm$0.12 & 0.52$\pm$0.10 & 0.56$^{+0.29}_{-0.31}$ & 1.54$\pm$0.40 \\
      Flare-3c (DEM) & 0.99$^{+0.04}_{-0.06}$ & 1.07$^{+0.10}_{-0.08}$ & 0.35$^{+0.08}_{-0.06}$ & 0.55$\pm$0.26 & 1.75$^{+0.44}_{-0.42}$ \\
      Flare-4a (2T) & 0.82$\pm$0.07 & 0.81$^{+0.05}_{-0.04}$ & 0.27$\pm$0.04 & 0.61$^{+0.11}_{-0.12}$ & 1.91$\pm$0.14 \\
      Flare-4a (DEM) & 0.65$\pm$0.06 & 0.69$^{+0.06}_{-0.04}$ & 0.21$\pm$0.04 &  0.71$\pm$0.12 & 2.29$\pm$0.16 \\
      Flare-4b (2T) & 1.31$^{+0.09}_{-0.08}$ & 1.40$^{+0.20}_{-0.18}$ & 0.58$^{+0.10}_{-0.11}$ & 1.04$^{+0.36}_{-0.33}$ &  1.50$^{+0.61}_{-0.60}$ \\
      Flare-4b (DEM) & 1.07$^{+0.08}_{-0.06}$ & 1.11$^{+0.16}_{-0.14}$ & 0.41$\pm$0.08 & 0.93$^{+0.30}_{-0.28}$ & 1.61$^{+0.62}_{-0.60}$ 
\enddata
\tablenotetext{a}{Relative to the solar photospheric values by \citet{Lodders2003}.}
\end{deluxetable}

\begin{thebibliography}{}
\bibitem[Allison et al.(2006)]{Allison2006}
  Allison, J., et al.\ 2006, IEEE Trans. Nucl. Sci. 53, 1, 270
\bibitem[Allison et al.(2013)]{Allison2013}
  Allison, J., et al.\ 2013, NIMA, 835, 186
\bibitem[Anders \& Grevesse(1989)]{Anders1989}
  Anders E., \& Grevesse N.\ 1989, Geochimica et Cosmochimica Acta, 53, 197
\bibitem[Apostolakis et al.(2003)]{Apostolakis2003}
  Apostolakis, J., et al.\ 2003, NIMA, 250
\bibitem[Argiroffi et al.(2004)]{Argiroffi2004}
  Argiroffi, C., et al. 2004, ApJ, 609, 925
\bibitem[Arnaud(1996)]{Arnaud1996}
  Arnaud, K.A.\ 1996, ASPC, 101, 17
\bibitem[Asplund et al.(2009)]{Asplund2009}
  Asplund M., Grevesse N., Sauval A.J., \& Scott P.\ 2009, ARAA, 47, 481
\bibitem[Audard et al.(2003)]{Audard2003}
  Audard, M., G\"udel, M., Sres, A., Raassen, A.J.J., \& Mewe, R.\ 2003, A\&A, 398, 1137
\bibitem[Baker et al.(2019)]{Baker2019}
  Baker, D., et al.\ 2019, ApJ, 875, 35
\bibitem[Brinkman et al.(2001)]{Brinkman2001}
  Brinkman, A.C., et al.\ 2001, A\&A, 365, L324
\bibitem[Brooks(2018)]{Brooks2018}
  Brooks, D.H.\ 2018, ApJ, 863, 140
\bibitem[Caspi et al.(2010)]{Caspi2010}
  Caspi, A., \& Lin, R.P.\ 2010, ApJ, 725, L161
\bibitem[Caspi et al.(2014)]{Caspi2014}
  Caspi, A., McTiernan, J.M., \& Warren, H.P.\ 2014, ApJL, 788, L31
\bibitem[Caspi et al.(2015)]{Caspi2015}
  Caspi, A., Woods, T.N., \& Warren, H.P.\ 2015, ApJL, 802, L2
\bibitem[Dennis et al.(2015)]{Dennis2015}
  Dennis, B.R., et al.\ 2015, ApJ, 803, 67
\bibitem[Doschek et al.(2015)]{Doschek2015}
  Doschek, G.A., Warren, H.P., \& Feldman, U.\ 2015, ApJL, 808, L7
\bibitem[Doschek \& Warren(2016)]{Doschek2016}
  Doschek, G.A., \& Warren, H.P.\ 2016, ApJ, 825, 36
\bibitem[Drake et al.(1997)]{Drake1997}
  Drake, J.J., Laming, J.M., \& Widing, K.G., 1997, ApJ, 478, 403
\bibitem[Favata et al.(2000)]{Favata2000}
  Favata, F., et al., 2000, A\&A, 353, 987
\bibitem[Feldman(1992)]{Feldman1992}
  Feldman, U. 1992, Phys.\ Scr.\ 1992, 46, 202
\bibitem[Feldman \& Widing(1993)]{Feldman1993}
  Feldman, U., \& Widing, K.G. 1993, ApJ, 414, 381
\bibitem[Fludra \& Schmelz(1999)]{Fludra1999}
  Fludra, A., \& Schmelz, J.T.\ 1999, A\&A, 348, 286
\bibitem[G\"udel et al.(2001)]{Gudel2001}
  G\"udel, M., et al. 2001, A\&A, 365, L336
\bibitem[Hedin(1991)]{Hedin1991}
  Hedin, A.E., 1991, J. Geophys. Res. 96, 1159
\bibitem[Henoux et al.(1998)]{Henoux1998}
  Henoux, J.-C.\ 1998, \ssr, 85, 215
\bibitem[Huenemoerder et al.(2003)]{Huenemoerder2003}
  Huenemoerder, D.P., Canizares, C.R., Drake, J.J., \& Sanz-Forcada, J. 2003, ApJ, 595, 1131
\bibitem[Huenemoerder et al.(2013)]{Huenemoerder2013}
  Huenemoerder, D.P., Phillips, K.J.H., Sylwester, J., \& Sylwester, B.\ 2013, ApJ, 768, 135
\bibitem[Ishisaki et al.(2007)]{Ishisaki2007}
  Ishisaki, Y., et al.\ 2007, PASJ, 59, S113
\bibitem[Itoh et al.(2002)]{Itoh2002}
  Itoh, M., et al. 2002, IAU Proceedings, 2, 435
\bibitem[Koyama et al.(2007)]{Koyama2007}
  Koyama, K., et al.\ 2007, PASJ, 59, S23
\bibitem[Labitzke et al.(1985)]{Labitzke1985}
  Labitzke, K., Barnett, J.J., \& Edwards B., 1985, Handbook MAP 16, SCOSTEP, University of Illinois, Urbana
\bibitem[Laming \& Drake(1999)]{Laming1999}
  Laming, J.M., \& Drake, J.J.\ 1999, ApJ, 516, 324
\bibitem[Laming(2004)]{Laming2004}
  Laming, J.M.\ 2004, ApJ, 614, 1063
\bibitem[Laming(2009)]{Laming2009}
  Laming, J.M.\ 2009, ApJ, 695, 954
\bibitem[Laming(2012)]{Laming2012}
  Laming, J.M.\ 2012, ApJ, 744, 115
\bibitem[Laming(2015)]{Laming2015}
  Laming, J.M.\ 2015, LRSP, 12, 2
\bibitem[Laming(2017)]{Laming2017}
  Laming, J.M.\ 2017, ApJ, 844, 153
\bibitem[Liefke et al.(2010)]{Liefke2010}
  Liefke, C., Fuhrmeister, B., \& Schmitt, J.H.M.M.\ 2010, A\&A, 514, A94
\bibitem[Lodders et al.(2003)]{Lodders2003}
  Lodders, K.\ 2003, ApJ, 591, 1220
\bibitem[Maeda et al.(2009)]{Maeda2009}
  Maeda, Y., et al.\ 2009, PASJ, 61, 1217
\bibitem[Meyer(1985)]{Meyer1985}
  Meyer, J.-P.\ 1985, ApJS, 57, 173 
\bibitem[Mitsuda et al.(2007)]{Mitsuda2007}
  Mitsuda, K., et al.\ 2007, PASJ, 59, S1
\bibitem[Nordon \& Behar(2008)]{Nordon2008}
  Nordon, R., \& Behar, E.\ 2008, A\&A, 482, 639
\bibitem[Narendranath et al.(2014)]{Narendranath2014}
  Narendranath, S., et al.\ 2014, \solphys, 289, 1585
\bibitem[Osten et al.(2003)]{Osten2003}
  Osten, R.A., et al.\ 2003, ApJ, 582, 1073
\bibitem[Pan et al.(1997)]{Pan1997}
  Pan, H.C., et al.\ 1997, MNRAS, 285, 735
\bibitem[Phillips \& Dennis(2012)]{Phillips2012}
  Phillips, K.J.H., \& Dennis, B.R.\ 2012, ApJ, 748, 52
\bibitem[Picone et al.(2002)]{Picone2002}
  Picone, J.M., Hedin, A.E., Drob, D.P., \& Aikin, A.C., 2002, J. Geophys. Res., 107(A12), 1468
\bibitem[Raymond et al.(1997)]{Raymond1997}
  Raymond, J.C., et al.\ 1997, \solphys, 175, 645
\bibitem[Reames(2018)]{Reames2018}
  Reames, D.V. 2018, SSR, 214, 61
\bibitem[Sanz-Forcada et al.(2003)]{Sanz-Forcada2003}
  Sanz-Forcada, J., Maggio, A., \& Micela, G. 2003, A\&A, 408, 1087
\bibitem[Schmelz et al.(2012)]{Schmelz2012}
  Schmelz, J.T., Reames, D.V., von Steiger, R., \& Basu, S.\ 2012, ApJ, 755, 33
\bibitem[Sterling et al.(1993)]{Sterling1993}
  Sterling, A.C., Doschek, G.A., \& Feldman, U.\ 1993, ApJ, 404, 394
\bibitem[Stern et al.(1992)]{Stern1992}
  Stern, R.A., Uchida, Y., Tsuneta, S., \& Nagase, F.\ 1992, ApJ, 400, 321
\bibitem[Sylwester et al.(1998)]{Sylwester1998}
  Sylwester, J., et al.\ 1998, ApJ, 501, 397
\bibitem[Sylwester et al.(2002)]{Sylwester2002}
  Sylwester, J., Kepa, A., \& Bentley, R.D.\ 2002, ASR, 30, 105
\bibitem[Sylwester et al.(2014)]{Sylwester2014}
  Sylwester, B., et al.\ 2014, ApJ, 787, 122
\bibitem[Sylwester et al.(2015)]{Sylwester2015}
  Sylwester, B., Phillips, K.J., Sylwester, J., \& Kepa, A.\ 2015, ApJ, 805, 49
\bibitem[Tanakashi et al.(2007)]{Takahashi2007}
  Takahashi, T., et al. 2007, PASJ, 59, S35
\bibitem[Tashiro et al.(2018)]{Tashiro2018}
  Tashiro, M., et al.\ 2018, Proc.\ SPIE, 10699, 1069922
\bibitem[Tawa et al.(2008)]{Tawa2008}
  Tawa, N., et al.\ 2008, PASJ, 60, S11
\bibitem[Tsuboi et al.(1998)]{Tsuboi1998}
  Tsuboi, Y., et al.\ 1998, ApJ, 503, 894
\bibitem[Tsuru et al.(1989)]{Tsuru1989}
  Tsuru, T., et al.\ 1989, PASJ, 41, 679
\bibitem[Veck \& Parkinson(1981)]{Veck1981}
  Veck, N. J. and Parkinson, J. H.\ 1981, MNRAS, 197, 41
\bibitem[Warren(2014)]{Warren2014}
  Warren, H.P.\ 2014, ApJL, 786, L2
\bibitem[Wood \& Linsky(2010)]{Wood2010}
  Wood, B.E., \& Linsky, J.L.\ 2010, ApJ, 717, 1279
\bibitem[Wood et al.(2018)]{Wood2018}
  Wood, B.E., Laming, J.M., Warren, H.P., \& Poppenhaeger, K.\ 2018, ApJ, 862, 66
\bibitem[ et al.()]{}


\end{thebibliography}
\end{document}